\documentclass[amsmath,amssymb,aps,pre,twocolumn,superscriptaddress]{revtex4-2}

\usepackage{latexsym}
\usepackage{amssymb}
\usepackage{mathrsfs}
\usepackage{amsmath}
\usepackage{soul}
\usepackage{bm}
\usepackage{verbatim}
\usepackage{epsfig}
\usepackage{multirow}
\usepackage{xspace}
\usepackage{ltablex}
\usepackage{ulem}
\usepackage[english]{babel}
\usepackage[utf8x]{inputenc}
\usepackage[T1]{fontenc}
\usepackage[usenames,dvipsnames]{xcolor}

\usepackage{color}
\definecolor{darkblue}{rgb}{0,0,0.6}
\definecolor{darkred}{rgb}{0.6,0,0}

\usepackage{hyperref}
\hypersetup{
	bookmarksopen=true,
	pdftitle="",
	pdfauthor="",
	pdfsubject="",
	pdfstartview={FitH},
	pdfmenubar=true,
	pdfhighlight=/O,
	colorlinks=true,
	urlcolor=darkblue,
	citecolor=darkblue,
	linkcolor=MidnightBlue,
}

\usepackage{cleveref}

\begin{document}

\title{The Role of Excitations in Supercooled Liquids: Density, Geometry, and Relaxation Dynamics}
 \author{Wencheng Ji}  
 \affiliation{Department of Physics of Complex Systems, Weizmann Institute of Science, Rehovot, 234 Hertzl St., Israel}
 \author{Massimo Pica Ciamarra} 
  \email{massimo@ntu.edu.sg}

 \affiliation{Division of Physics and Applied Physics, School of Physical and Mathematical Sciences, Nanyang Technological University, 21 Nanyang Link, 637371, Singapore}
\affiliation{Consiglio Nazionale delle Ricerce, CNR-SPIN, Napoli, I-80126, Italy}
 \author{Matthieu Wyart}
  \email{matthieu.wyart@epfl.ch}
\affiliation{Institute of Physics, \'Ecole Polytechnique Fédérale de Lausanne, Lausanne, CH-1015, Switzerland}


\renewcommand{\floatpagefraction}{.5}

\begin{abstract}
Low-energy excitations play a key role in all condensed-matter systems, yet there is limited understanding of their nature in glasses, where they correspond to local rearrangements of  groups of particles. Here we introduce an algorithm to systematically uncover these excitations up to the activation energy scale relevant to structural relaxation. We use it in a model system to  measure the density of states on a scale never achieved before, confirming that this quantity shifts to higher energy under cooling, precisely as the activation energy does. Secondly, we show  that the excitations' energetic and spatial features allow one to predict with great accuracy   the dynamic propensity, i.e. the location of future  relaxation dynamics. Finally, we find that  excitations have a core whose properties, including the displacement of the most mobile particle, scale as a power-law of their activation energy and are independent of temperature. Additionally, they exhibit an outer deformation field that depends on the material's stability and, therefore, on temperature. 
We build a scaling description of these findings. Overall, our  analysis supports that excitations play a crucial role in regulating relaxation dynamics near the glass transition, effectively suppressing the transition to dynamical arrest predicted by mean-field theories while also being strongly influenced by it.

\end{abstract}


\maketitle

\section*{Introduction}\label{sec1}
In low-temperature glasses, elementary excitations are two-level systems, groups of particles that can tunnel between two states~\cite{Phillips72,Anderson72,Phillips87,Queen13,Perez14,Khomenko20}.
At much higher temperatures near the glass transition $T_g$,  structural relaxation occurs on a time scale $\tau=\tau_0 \exp(E_a/T)$, where $\tau_0$ is microscopic time and $E_a$ is an activation energy that grows on cooling in fragile liquids~\cite{angell1985strong}. 
Thermally activated elementary rearrangements of a few particles are important in this temperature range as well, as they contribute to structural relaxation \cite{Simmons2012, Ciamarra2015SM, Cicerone2014, cicerone2023excitation}. 
Such rearrangements thus play a vital role in theories of the glass transition ~\cite{Anderson95, Debenedetti01, dyre2023solid, Berthier2018}.
In Kinetically Constrained Models, local rearrangements are defects that can diffuse and interact to relax the system~\cite{garrahan2002geometrical, fraggedakis2023inherent}. 
In mean field theories, local rearrangements correspond to the string-like~\cite{gotze1988scaling, schweizer2003entropic, vollmayr2002dynamical, charbonneau2014hopping} `hopping processes' through which finite-dimensional systems relax below the dynamical or mode coupling temperature $T_c$, where the dynamics would halt in infinite dimensions where these processes are absent~\cite{Lubchenko01,Biroli12}.
Finally, in elastic models of the glass transition \cite{dyre2006colloquium,rainone2020pinching,JeppeEdan,Li2022, Lerbinger2022}, the energy of local excitations directly determines the activation energy $E_a$.
Theoretically, distinct ideas have been proposed to understand the geometry of local rearrangements, including entropic considerations \cite{Stevenson10} or the existence of defects around a hexatic phase in two dimensions \cite{fraggedakis2023inherent}.
Alternatively, building on the notion of a length scale diverging~\cite{franz2000non, Biroli06, franz2011field}  at the dynamical transition $T_c$, Ref~\cite{Wencheng22} derived relationships between geometric and energetic properties of the excitations with the minimal energy.

Differentiating between different scenarios for the geometry of these excitations requires measuring their density of states $N(E)$ and geometry across a broad energy spectrum, a challenge that remains unresolved.
Indeed, potential energy landscape studies \cite{Doliwa2003,Heuer08} access the consecutive excitations activated in a liquid during its relations but cannot provide the density of states of excitations. 
Studies conducted on the few lowest-energy excitations \cite{Wencheng19, Schober93, Wang19, Khomenko20, Wencheng20}, are pertinent to the plastic and quantum properties of glasses and not directly relevant to the glass transition, which typically involves much higher energy rearrangements.
We recently developed SEER~\cite{Massimo23}, an algorithm based on thermal exploration that allowed us to measure $N(E)$.
However, SEER only accesses excitations with energy notably smaller than the activation energy $E_a=T \log(\tau/\tau_0)$ and hence does not inform on many excitations that have a significant probability to be triggered dynamically and contribute to structural relaxation~\cite{Doliwa2003}.
The excitations' density of state and geometric properties in such a broad energy range have yet to be studied, as well as the connection between these observables and the dynamics.

\begin{figure*}[htb]%
\centering
\includegraphics[width=0.6\linewidth]{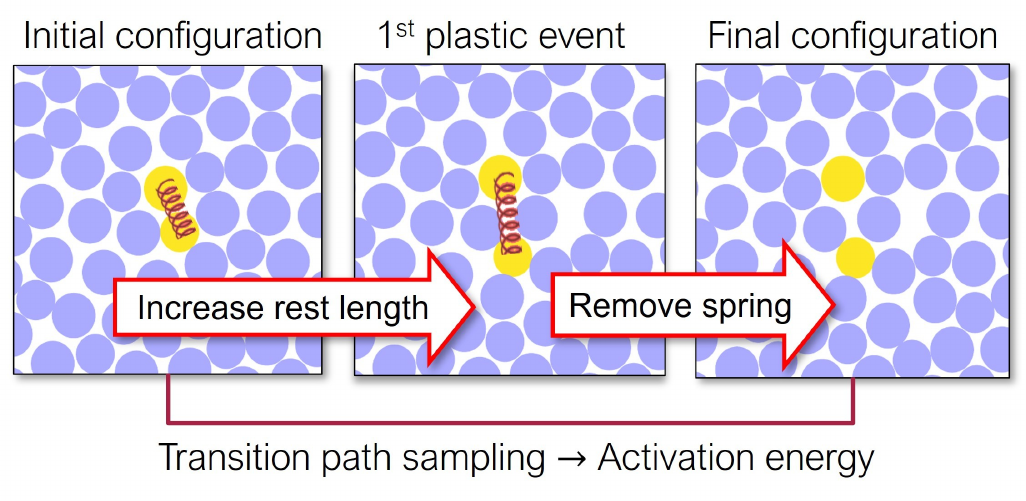}
\caption{Illustration of the ASEER algorithm. 
We detect excitations by enforcing a dipolar force through a spring connecting adjacent particles of energy-minimized configurations. 
We gradually increase the spring rest length until an instability occurs. We then remove the spring and minimize the energy, bringing the system to a novel energy minimum. 
The minimum energy pathway going from one state to the other is analyzed in the absence of the spring.} 
\label{fig:sketch}
\end{figure*}

Indeed, the hypothesis that excitations govern where future structural relaxation will occur remains to be tested. 
A key quantity to predict is the particle propensity~\cite{WidmerCooper2004}, which is the squared particle displacement averaged over an iso-configurational ensemble of trajectories. 
These trajectories share the same initial configuration but differ in their momenta, which are randomly drawn from the expected equilibrium distribution. 
If excitations regulate the relaxation dynamics, they should enable the prediction of propensity at later times, with high-energy excitations becoming increasingly relevant for these predictions over time.

In this work, we introduce an algorithm to identify excitations via mechanical perturbations.
This algorithm offers fast computation and enables access to a wide range of energies, surpassing the activation energy of the model under consideration. 
Our findings reveal three key insights:
(i) We observe that the density of states $N(E)$ approximately follows $N(E)\propto (E-E_g(T))^a$ up to the activation energy. Additionally, we confirm that the variation in $E_g(T)$ predicts the change in activation energy \cite{Massimo23}.
(ii) We introduce an excitation-based propensity predictor that successfully correlates with the propensity and demonstrate that high-energy excitations contribute to long-term relaxation.
(iii) Remarkably, we show that certain excitations' geometric properties, such as the displacement $\delta$ of the most-moving particle or the probability of displaying string-like motion, only depend on their energy $E$. However, other properties, like the volume of an excitation, depend on both energy and temperature.
We reconcile these observations by considering that excitations possess a core solely governed by their energy, which then influences the surrounding medium on a scale determined by material stability and, consequently, temperature.
Overall, our results suggest that the hopping processes that suppress the mean-field dynamical transition are very much affected by it.

\section*{ASEER:  Athermal Systematic Excitation ExtRaction }\label{subsec:seer}
We propose an athermal algorithm, the Athermal Systematic Excitation ExtRaction or `ASEER', to uncover the excitations associated with a reference inherent structure (IS$_0$).  
This protocol builds on the idea that changing the topology of the Voronoi neighbors induces an excitation, in analogy with the triggering of T1 transition in two spatial dimensions, see, e.g., Ref.~\cite{Li2022,Lerbinger2022,Popovi2021}. 
To uncover an excitation, we increase the separation between two adjacent (\`a la Voronoi) particles $i$ and $j$ by modifying the energy functional via the addition of an elastic spring, $E_{\rm spring}(\Delta r)=k[(r_{ij}^0+\Delta r)-r_{ij}]^2$, $r_{ij}^0$ being the distance between the particles in IS$_0$. 
We have checked that the value of $k$ is not critical and empirically fix it so that the spring is never compressed by more than $5\%$.
We slowly increase $\Delta r$ while continuously minimizing the energy to keep the system in a minimum of the expanded energy functional $U(\{{\bf r_i}\})+E_{\rm spring}(\Delta r)$.  
The $\Delta r$ dependence of the spring energy (or of the total one) comprises smooth elastic branches punctuated by sudden drops corresponding to irreversible rearrangements. 
We focus on the first plastic event identified via a standard thresholding approach.
When this event occurs, we remove the spring and minimize the energy again, potentially leading the system to a new IS$^{*}$.
This approach is illustrated in Fig.\ref{fig:sketch} and involves constraining one degree of freedom.
We investigate the minimum energy path connecting IS$_0$, and each uncovered IS$^*$ via the nudge-elastic-band (NEB) method~\cite{neb1}
to estimate the energy $E_{\rm Saddle}$ of the saddle point separating the considered ISs.
If not stated otherwise, when the minimum energy path traverses additional ISs ($\simeq 20\%$ of cases in the considered system), we redefine IS$^*$ as the first encountered IS and repeat the NEB analysis.
This results in a catalog of unique (we ensure each IS$^{*}$ appears once) excitations, each characterized by its energy barrier $E=E_{\rm Saddle}-E_{\rm IS_0}$ and displacement field $\bf{dr} = \bf{r^*}-\bf{r}^0$, with $\bf{r^*}$ and $\bf{r}^0$ the positions in the two ISs.
This algorithm leads to large catalogs of excitations, as it uncovers from $1.2$ to $2$ excitations per particle on cooling. 
As demonstrated below, at low energies, ASEER matches previous algorithms systematically searching excitations via thermal fluctuations, while it performs much better at energies of the order of the activation energy $E_a$.

\section*{Density of states}
\begin{figure}[t!]
\centering\includegraphics[width=1\linewidth]{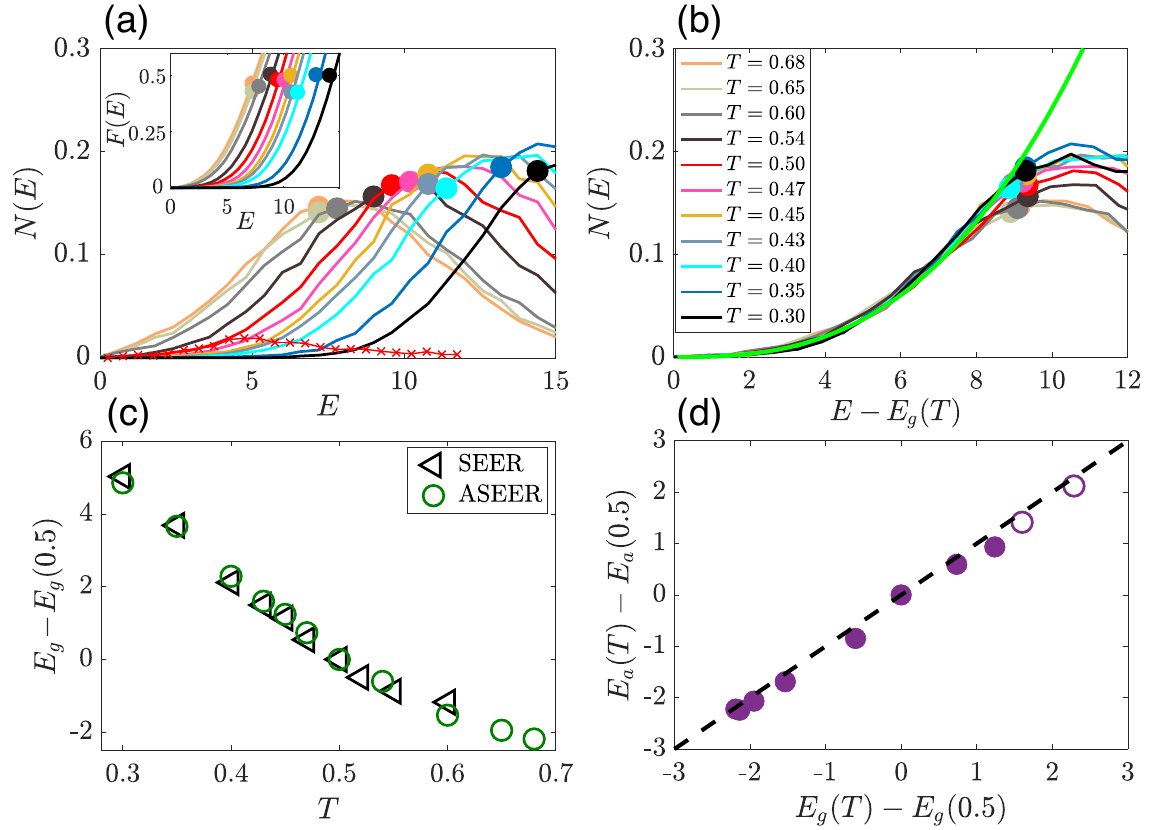}
\caption{
(a) Density of excitations $N(E)$,  normalized by the system size $\mathcal{N}$, and its cumulative distribution $F(E)$ (inset).
The solid dots mark the values of the activation energy~\cite{Massimo23}, $E_a(T)=T \log\left(\tau/\tau_0\right)$, and the cross-shaped line has been obtained with the SEER algorithm at $T=0.5$.
(b) The $N(E)$ curves collapse when the energy is shifted by $E_g(T)$. 
The thick green curves is $N(E)\!\approx g_1\!\times\!(E\!-\! E_g)^{2.7\pm0.1}$, where $g_1\!=\!\!(4.5\pm 0.5)\!\times\! 10^{-4}$.
(c) The shift of $E_g$ on cooling matches the shift in activation energy evaluated via SEER \cite{Massimo23}. 
(d) The increase in $E_g$ matches the increase in activation energy measured from the relaxation dynamics, with $T=0.5$ an arbitrary reference temperature. 
Open circles correspond to the low-temperature values at which we use the time-temperature superposition to estimate the relaxation time and $E_a$.
} 
\label{fig:N_E}
\end{figure}

We applied ASEER to a polydisperse three-dimensional system of $N=2000$ soft repulsive particles~\cite{Lerner19} that can be equilibrated up to experimentally comparable temperatures through the `swap' algorithm \cite{Glandt84,gutierrez2015static,Ninarello17,Brito18}.
In recent work~\cite{Massimo23}, we determined this model's relaxation time $\tau(T)$ and microscopic timescale $\tau_0$, and hence the temperature dependence of its activation energy, $E_a(T) = T\log(\tau/\tau_0)$. For the two lowest temperatures (open circle), we estimate $\tau$ through the time-temperature superposition principle~\cite{Berthier2020,SM,Massimo23}.
In the Materials and Methods Section, we provide numerical details on the numerical model and the measure of the relaxation time.

Fig.~\ref{fig:N_E}(a) illustrates the energy $E$ dependence of the density of excitations $N(E)$, 
and its cumulative $F(E)$(inset).
At each temperature $T$, we average over ten samples. 
Notably, $N(E)$ and  $F(E)$ approximately shifts towards higher energy values as $T$ decreases: the energy of local barriers grows under cooling.  
Indeed, before its maximum $N(E)$ is well-described by:
\begin{equation}
N(E)\propto (E\!-\!E_g(T))^a.
\end{equation}
with $a\approx 2.7$, as demonstrated by the data collapse in Fig.~\ref{fig:N_E}(b).

The energy gap $E_g$ characterizing the system's stability increases on cooling, as shown in Fig.~\ref{fig:N_E}(c).
Fig.~\ref{fig:N_E}(d) shows that the variation of $E_g$ on cooling is consistent with that of the activation energy, indicating that local barriers control the dynamics in this liquid. 
The increase of a gap $E_g$ is reminiscent of the mean-field prediction according to which, for $T < T_c$, the density of vibrational modes is gapped up to a frequency $\omega_{\min}$ indeed increasing on cooling \cite{cavagnaSGpedestrians}.

In Fig.~\ref{fig:N_E}(a) and (c), we also present data obtained previously using the SEER algorithm \cite{Massimo23}. 
While both algorithms yield consistent results for the variation of $E_g$ (and of other quantities in Fig.~S1 of \cite{SM}), ASEER notably detects excitation on the scale $E_a$, which can be activated on the relaxation time scale.
Each algorithm possesses distinct advantages: SEER, reliant on thermal cycles, uncovers excitations with the smallest $E$, making it ideal for obtaining accurate statistics at lower $E$ values where $N(E)$ is limited. 
Conversely, the mechanical ASEER algorithm accesses a wide spectrum of activation energies.
Noticing that the algorithms give consistent results in the energy range accessed by both (Fig.~S1),
we integrate them in the subsequent analysis to acquire high-quality statistics across a broad energy range.

\section*{Excitations predict the spatiotemporal relaxation dynamics} 
\begin{figure}[!!t]%
\centering
\includegraphics[scale=0.6]{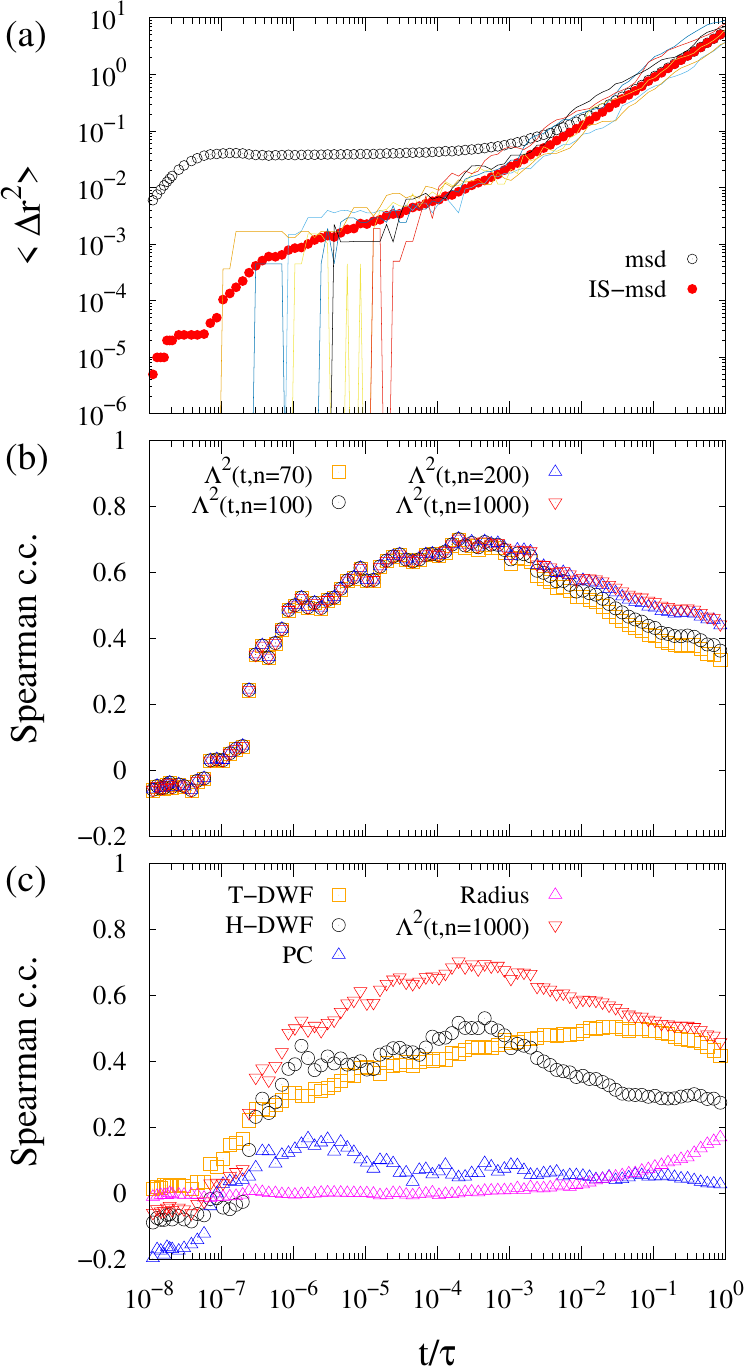}
\caption{
(a) Symbols identify the finite-temperature and IS mean square displacements averaged over configurations and iso-configurational runs. Lines are examples of IS mean square displacement of individual trajectories.
(b) Time dependence of the Spearman correlation coefficient between the particle propensity and the excitation-based predictor, for varying number of considered excitations. SEER extract $n\approx70$ for each configuration. ASEER many more, and $n\approx1000$ are those with activation energy smaller than $T \log(\tau/t_0)$.
(c) Time dependence of the Spearman correlation coefficient between the particle propensity and various predictors: the Debye Waller factor (T-DWF) and its Harmonic approximation (H-DWF), the packing capability (PC), and the excitation-based predictor. All data refer to $T=0.45$, below the estimated mode-coupling critical temperature for this model, $T_c \in [0.51:0.54]$.} 
\label{fig:propensity}
\end{figure}

To demonstrate the relevance of excitations to the relaxation dynamics further, we examine their ability to predict structural relaxation. 
As a measure of structural relaxation, we focus on the propensity of motion~\cite{WidmerCooper2004}.
While particles with a high propensity are interpreted as more prone to structural relaxation, a particle's propensity also depends on its vibrational motion.
Indeed, by comparing the standard and the inherent structure mean square displacements, both averaged over configurations and iso-configurational trajectories,
Fig.~\ref{fig:propensity}(a) demonstrates that the vibrational contribution dominates the mean square displacements for a long transient.
To ensure the propensity informs on structural relaxation, here we filter out the vibrational contribution by defining it from the inherent structure mean square displacement instead of the previously used finite temperature displacement.
Henceforth, the propensity of particle $i$ at time $t$ is $p_i(t)=\langle {\bf \Delta r}_{i,\rm IS}^2(t)\rangle_{\rm isoconf}$, with ${\bf \Delta r}_{i,\rm IS}(t) = {\bf r}_{i,\rm IS}(t)-{\bf r}_{i,\rm IS}(0)$, and ${\bf r}_{i,\rm IS}(t)$ the position of the particle in the IS associated with the configuration visited by the system at time $t$.

We define a predictor \(\Lambda_{i}^2(t,n)\) for the propensity of motion by analyzing the \(n\) excitations with the smallest activation energy associated with the initial configuration.
To this end, we treat an excitation as a double potential well and consider it inactive when the system is in the minima of the $t=0$ configuration; it is active otherwise. 
Our proposed excitation-based predictor is
\begin{equation}
\Lambda_i^2(t,T_p,n) = \sum_{k=1}^{n} {\bf u}_{i,k}^2 P_{k}(T_p,t),
\end{equation}
where \({\bf u}_{i,k}\) is the displacement of particle \(i\) in excitation \(k\).
Here 
$P_k(T_p,t) = P_{\rm eq} \left[ 1 - e^{-(\Gamma_f+\Gamma_b)t}\right]$ 
is the probability that the excitation is active at time \(t\) within the approximation that excitations do not interact, 
$\Gamma_{f,b} = t_0^{-1} e^{-\Delta E_{f,b}/T_p}$ are the forward and backward transitions rates, with $\Delta E_{f,b}$ the corresponding activation energies,
and $P_{k,\rm eq} = \frac{\Gamma_f}{\Gamma_f+\Gamma_b}$ is the equilibrium activation probability.
We average the results over 5 independent initial configurations and consider 20 iso-configurational trajectories for each configuration.

Figure~\ref{fig:propensity}(b) illustrates the time dependence of the Spearman's rank correlation coefficient between \(\Lambda_i^2(t,n)\) and the IS-propensity \(p_{i}(t)\) at $T=0.45$.
We consider $n\approx70$, the number of excitations per configuration typically extracted by the SEER algorithm~\cite{Massimo23}.
Other considered $n$ values are bounded by the number of excitations with activation energy smaller than the activation energy $E_a$ regulating structural relaxation, which is 
$n\approx1000$ according to Fig.~\ref{fig:N_E}(b).
We find the correlation coefficient is $n$ independent up to its maximum value of $\simeq 0.7$, which occurs at $t \simeq 10^{-3}\tau$, when the mean square displacement crossovers towards the diffusive regime.
This result demonstrates that only low-energy excitations contribute to the short-time dynamics.
Subsequently, the correlation coefficient decreases.
It decreases faster for smaller $n$ values, showing that at later times, high-energy excitations contribute to the relaxation process.

Previous works have performed related studies by considering predictors defined by analysing the structural and elastic properties of the initial configuration, see~\cite{Tanaka2019} for a review.
In the following, we compare our excitation-based predictor to physically motivated (as opposed to machine-learning-derived~\cite{jung2023roadmap}) ones:
(i) The Debye-Waller factor T-DFW = $\langle \Delta r^2_i(t_{\textsc{DWF}}) \rangle$, with $t_{\textsc{DWF}}$ the estimated \textsc{DWF} time~\cite{widmer2008irreversible};
(ii) The Debye-Waller factor evaluated in the harmonic approximation H-DFW = $\sum_k {\bf e}_{i,k}^2\omega_k^{-2}$, where the sum is over the modes of the dynamical matrix associated with the $t=0$ inherent structure, ${\bf e}_{i,k}$ is the displacement of particle $i$ in mode $k$, and $\omega_k$ the mode eigenfrequency~\cite{Tong2014};
(iii) The local packing capability (PC), which measures how well the particles surrounding the considered one are well packed~\cite{Tong2019}.
(iv) The particle radius.
We highlight our approach is the only one whose predictions have a time dependence.

Fig.~\ref{fig:propensity}(c) demonstrates that the excitation-based predictor outperforms all other predictors, except in the very short time ballistic regime. 
The finite temperature Debye-Waller factor demonstrates a similar predictive ability close to the relaxation time.
The Harmonic DWF has a poorer predictive ability, confirming the absence of strong correlations between the local elastic properties in the linear response regime and the relaxation dynamics~\cite{Li2022,Patinet2022}.
The packing capability and particle radius have the worst predictive ability, with the radius's predictive power trivially increasing over time since smaller particles diffuse more than larger ones. This suggests that polydispersity disrupts the correlation between local geometrical properties and structural relaxation observed in mono- and bi-disperse systems. 

\section*{Architecture of excitations} 

{\it Architecture of lowest-energy excitations --} 
\begin{figure}[t!]
\centering
\includegraphics[width=1\linewidth]{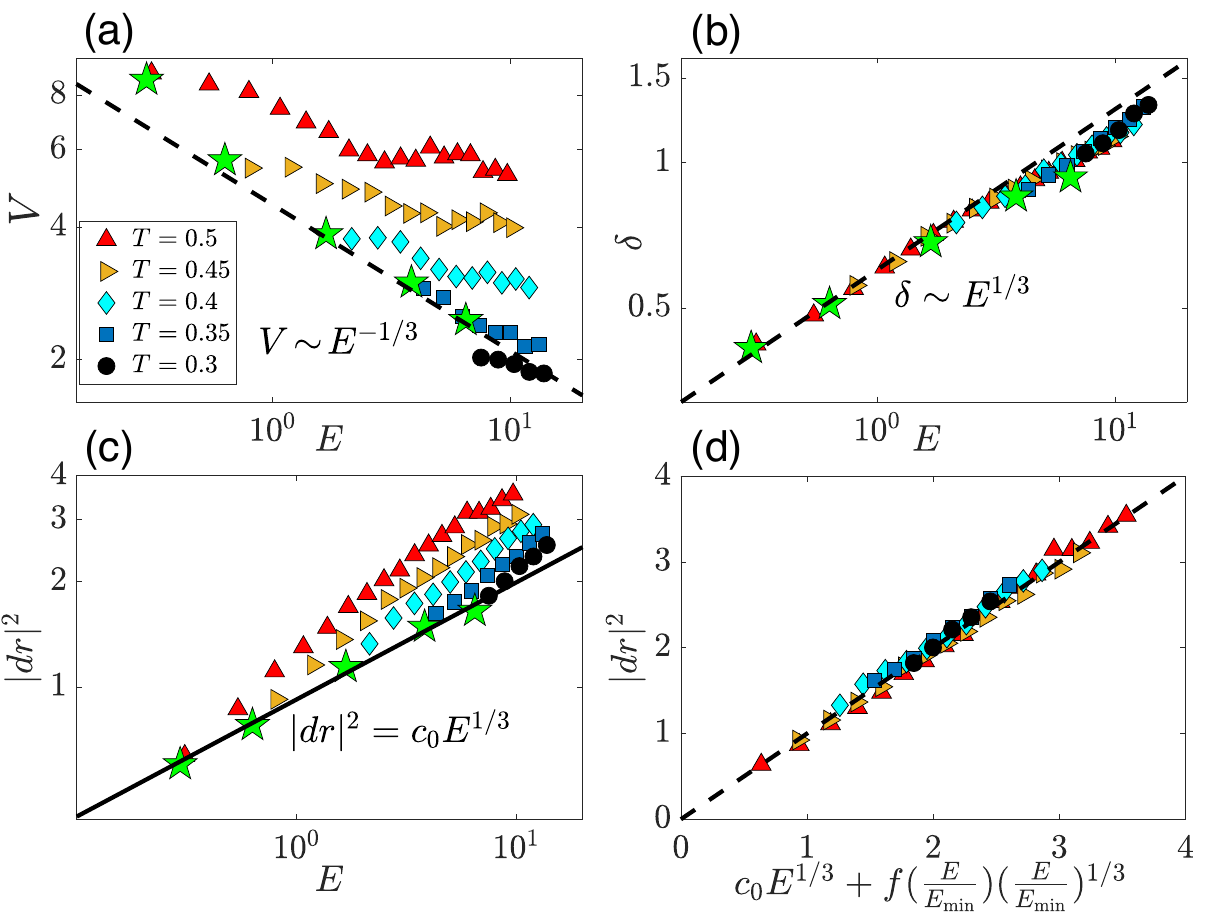}
\caption{
Dependence of (a) volume, (b) largest particle displacement, and (c) squared norm of the displacement field on the excitation's activation energy, for various temperatures.
The green stars refer to the excitations with the smallest activation energy. In all panels, we plot median values (see Fig.~S2 in \cite{SM} for mean values).
(d) The data in (c) are collapsed by a model postulating that the excitation core induces a $T$-dependent far-field displacement. 
 }
\label{fig:core}
\end{figure}
We analyze the geometrical properties of each excitation by its displacement field, ${\bf dr}$, focusing on:
(i) the characteristic number of involved particles, estimated by the participation ratio $V\equiv NP_r = (\sum {\bf dr}_i^2)^2/(\sum {\bf dr}_i^4)$, where the sums run over all particles $i$;  (ii) the squared norm of the displacement field, $|dr|^2 \!=\!\sum_i {\bf dr}_i^2$; (iii) the maximum particle displacement, $\delta=\max_i ||{\bf dr}_i||$.

By elaborating on the mean-field scenario~\cite{franz2000non, Biroli06, franz2011field}, Ref.~\cite{Wencheng22} predicted that for the excitations with minimal energy, associated with modes near the gap:
\begin{equation}
     V_{\min}(T)\sim \frac{1}{\delta_{\min}(T)}\sim \frac{1}{ |dr_{\min}(T)|^2} \sim \frac{1}{{E_{\min}(T)}^{\frac{1}{3}}}.
     \label{E_scale}
 \end{equation}
These excitations are also associated with a length scale $\ell_{\min}(T)\sim \sqrt {V_{\min}(T)}$, which is empirically known to characterize the linear response to an imposed dipole ~\cite{lerner14,Rainone20}. 
Ref. \cite{Wencheng22} verified the relationships between $V_{\min}$, $\delta_{\min}(T)$ and $|dr_{\min}(T)|^2$, but the activation energy $E_{\min}$ was not measured; instead a proxy corresponding to the energy difference between the two IS was used.
We validate the scaling of the excitation architecture with $E_{\min}$ in Fig.~\ref{fig:core}(a), (b) and (c) (green stars) by investigating, at each temperature, the features of the lowest-energy excitation of 1000 systems.

Note that we exclude string-like excitations from the geometrical analysis (they are defined as excitations involving the exchange of particles; see below). In general, including them only affects results for the largest energies considered, as shown in Fig. S4~\cite{SM}.

\begin{figure}[t!]
\centering
\includegraphics[width=0.8\linewidth]{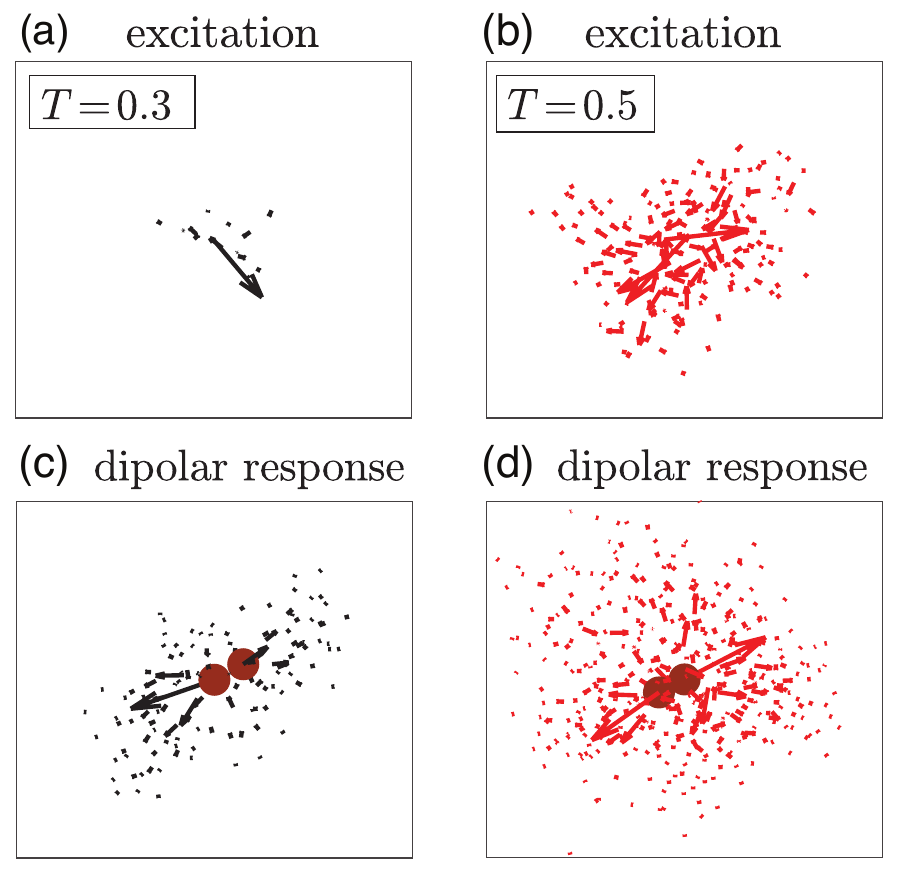}
\caption{Representative 2D-projections of the displacement fields of excitations $E\approx 10$ and $\delta\approx 1$ at temperatures $T=0.3$ (a) and $T=0.5$ (b). The displacement fields are scaled by a factor of three for visualization purposes.
Response fields induced by a dipole force acting on neighboring particles (brown dots) at $T=0.3$ (c) and $T=0.5$ (d).
\label{fig:dis_example}}
\end{figure}

{\it Architecture of high-energy excitations --}
Figure~\ref{fig:core} illustrates our main results, how the excitations' geometric properties depend on $E$ and $T$, for $E>E_{\min}$. 
Panel (b) shows a remarkable result: the largest displacement in an excitation depends on the excitation energy but not on its temperature so that $\delta \propto E^{1/3}$. 
By contrast, panels (a) and (c) demonstrate that, at fixed energy, the volume of an excitation and its squared norm increase with temperature. 
These facts can be visualized by considering the two-dimensional projection of the displacement field of excitations of similar energy $E\approx 10$ in systems having different temperatures, $T=0.3$ and $T=0.5$. 
Fig.~\ref{fig:dis_example} (a) and (b) show that these excitations have a comparable maximal magnitude at their center, but differ in the far field.

We rationalize these observations by assuming that excitations consist of (i) a core that depends only on energy $E$ and (ii) a far-field elastic response triggered by the core. In this view, the norm square of an excitation consists of two parts $|dr|^2(T,E) = |dr|^2_{\rm core}+|dr|^2_{\rm{f.f.}}$, which we now estimate. 
 
Assumption (i) implies that the core is the same if $E=E_{\min}$ or $E>E_{\min}$, as long as the energy is the same. Together with Eq.~\ref{E_scale}, it thus leads to a maximal displacement of the core, and therefore of the whole excitation, behaving as $\delta \sim E^{1/3}$ and a square norm of the core $|dr|^2_{\rm core}(E) \propto \delta^2(E)V(E) = c_0 E^{1/3}$.

If $E\gg E_{\min}$, we expect this core to act as a local strain, which generically triggers a dipolar linear elastic response in the far field  \cite{Picard2005}.
It is known that the response to a unit dipole is more extended at large rather than small temperatures~\cite{lerner14,Rainone20}, as we explicitly show in Fig.~\ref{fig:dis_example} (c) and (d). Quantitatively, 
following \cite{lerner14,Wencheng22} we expect the volume of a unit dipole response to vary with temperature as $V_{\min}(T)$, of order $\sim E_{\min}(T)^{-1/3}$ according to Eq.~\ref{E_scale}. Moreover, we expect the norm of the dipole response to be proportional to the norm of the core itself. Putting these two effects together, we obtain  $|dr|^2_{\rm f.f.} \propto f(E/E_{\min})|dr|^2_{\rm core}(E)V_{\min}(T) \propto f(E/E_{\min}) (E/E_{\min})^{1/3}$.
Here the function $f(x)$, with $f(1)=0$ and $\lim_{x\rightarrow\infty} f(x)= C>0$, accounts for the fact that only high-energy excitations ($E\gg E_{\min}$) produce a far-field response that differs from the core.
Overall, we find:
\begin{eqnarray}
|dr|^2(T,E) = c_0E^{1/3}\!+\!f\!\left(\!\frac{E}{E_{\min}}\!\right)\left(\frac{E}{E_{\min}}\right)^{1/3}.
\label{dr2}
\end{eqnarray}
Fig.~\ref{fig:core}(d) validates this theoretical prediction for a natural choice of interpolating function $f(x)\propto (x-1)/(x+1)$. In Fig.~S3 of SM, we show that Eq.~\ref{dr2} also holds if $E_{\min}$ is replaced by $E_{5}$, $E_{10}$ (respectively the median of the five or ten lowest energy excitation) or $E_g$. 

\begin{figure*}[htb]
\centering
\includegraphics[width=0.8\linewidth]{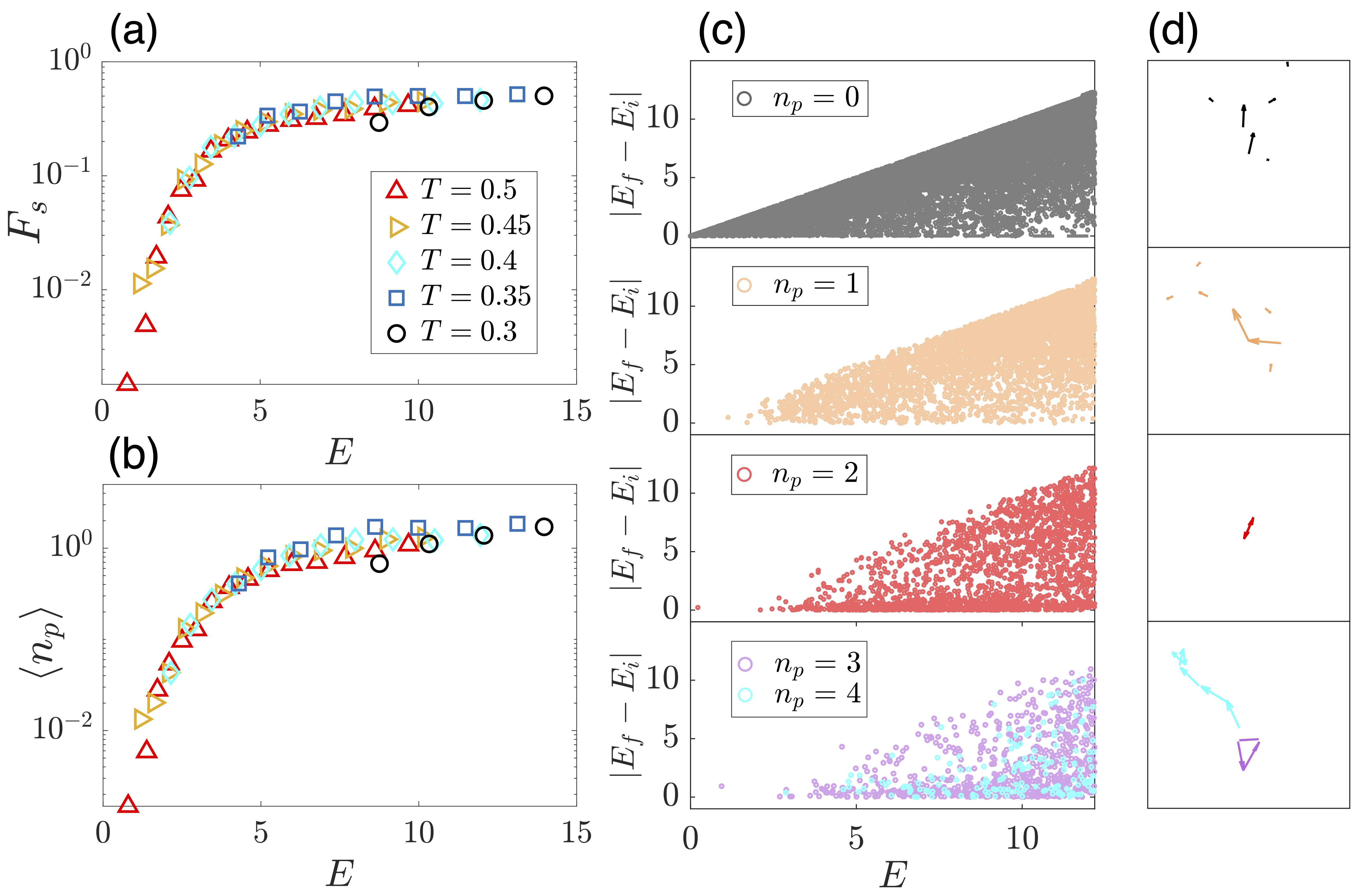}
\caption{
Energy dependence of (a) the fraction of string-like excitations and (b) the average string length, $n_p$. (c) Scatter plots of the energy difference between the final and initial inherent structures (ISs) versus $E$, for various $n_p$. (d) Projection of displacement fields for representative excitations onto a 2D plane.
We illustrate the emergence of a string as the system moves along its minimum energy path in Animation 1~\cite{SM}.
\label{fig:strings}
}
\end{figure*}

{\it String-like excitations (SLEs) -- }
Numerical simulations~\cite{Donati98,Yu17,Wencheng20} and recent experiments~\cite{Chen2023} show string-like particle motion in low-temperature structural relaxation.
String-like rearrangements involve one or more particles swapping positions in the excitation core, with displacements comparable to inter-particle separation~\cite{Wencheng22}. 
We thus expect that
(i) High-energy excitations are more likely string-like due to their larger displacements.
(ii) The probability of a string-like excitation is $T$-independent as the core properties are $T$-independent.

We identify SLEs by assuming that particle $i$ ends up in the position originally occupied by particle $j\neq i$ if $|{\bf{r}}^*_j\!-\! {\bf{r}}_i^0| < \Delta$, ${\bf{r}}_i^0$ and ${\bf{r}}^*_j$ being the positions in the initial and final configurations.
The choice of $\Delta$ is not critical as long as its value corresponds to a small fraction of the inter-particle distance. Here, we fix $\Delta=0.1$.
The number of particles involved in a string defines a string length
$ n_p = \sum_i\sum_{j,j\neq i}\theta\left(0.1-|{\bf{r}}^*_j\!-\!  {\bf{r}}_i^0|\right)$.
An excitation is string-like provided $n_p \geq 1$. An example string-like excitation is in Animation 1~\cite{SM}.
Fig.~(\ref{fig:strings})(a) shows that an excitation is a string with a $T$-independent probability $F_s=\langle \theta(n_p-1)\rangle$ that increases with $E$, consistently with our expectations.
The string length similarly increases with $E$ and is approximately $T$-independent, as we illustrate in Fig.~(\ref{fig:strings})(b).

We delve deeper into the dependence of an excitation's spatial and energetic properties on the string length. 
Figure (\ref{fig:strings})(c) demonstrates how the energy change $|E_f-E_i|$ between the initial and final inherent structures varies with $E$, considering strings of different lengths at $T=0.4$. 
Notably, only a few excitations with $n_p=0,1$ exhibit $|E_f-E_i| \simeq 0$, while this trend becomes more prevalent for larger $n_p$, particularly for $n_p = 2$. 
This phenomenon arises because many excitations with greater $n_p$ correspond to closed strings, as depicted in Fig. (\ref{fig:strings})(d), and consequently have minimal impact on the system's structure and energy. 
By promoting particle motion without inducing substantial structural relaxation, these closed strings may also contribute to the breakdown of the Stokes-Einstein relation in deeply supercooled liquids, in addition to dynamical heterogeneities \cite{ediger2000spatially}.

\section*{Conclusion}\label{sec13}
Through the novel ASEER algorithm, we have measured the density of excitations in a model glass-former up to the activation energy.
We confirmed that a shift in this density under cooling can predict the fragility of the liquid~\cite{Massimo23} and demonstrated that the excitations successfully predict the spatiotemporal relaxation dynamics, in particular they predict the particle propensity with accuracy.
Most importantly, we have demonstrated that the geometry of excitations depends on both their energy and temperature, which in turn governs the stability of the overall medium.
Excitations display a core whose properties scale with their energy and a far field component whose length scale is governed by temperature.
These insights align with known observations on the geometry of relaxation in liquids, such as the increased predominance of strings under cooling.

These results support the idea that excitations are influenced by a dynamical transition. 
On one hand, the shift in the density of excitations mirrors the shift in the Hessian of the energy landscape predicted in high dimensions. 
This shift differs from a simple rescaling, as one might expect from a naive interpretation of elastic models where energies are scaled by a temperature-dependent elastic modulus. 
On the other hand, the excitation core follows scaling laws expected in the vicinity of a dynamical transition. 
According to this perspective, hopping processes suppress the divergence of the relaxation time at $T_c$ but are very much influenced by the elastic instability associated with $T_c$.
\\
\\
{\bf Acknowledgments:} 	We thank E. Lerner for providing some of the equilibrium configurations we investigated. We thank the Simons collaboration as well as L. Berthier, G. Biroli, C. Brito, C. Gavazzoni, E. Lerner,  M. Muller, M. Popovic, M. Ozawa and A. Tahaei for discussions.
M.P.C.  discloses support for the research of this work from Singapore Ministry of Education under grants MOE-T2EP50221-0016 and T1RG152/23. M.W acknowledges support from the Simons Foundation
Grant (No. 454953 Matthieu Wyart) and from the
SNSF under Grant No. 200021-165509.

\bibliography{new}
\bibliographystyle{apsrev4-1}

\onecolumngrid

\begin{center}
	\rule{200pt}{0.5pt}
\end{center}

\appendix
\newpage

\clearpage

 \begin{center}
 	\medskip
 	\textbf{Appendix}
 \end{center}

\setcounter{figure}{0}
\renewcommand{\thefigure}{S\arabic{figure}}

\subsection*{Numerical model, relaxation dynamics and activation energy} 
We consider a three-dimensional system of soft repulsive particles~\cite{Lerner19} with size $\sigma$ distributed as $p(\sigma) \propto \sigma^{-3}$ in the range [$\sigma_{\rm min}$:$2.2\sigma_{\rm min}$].
	This modern numerical model can be equilibrated up to experimentally comparable temperatures through the `swap' algorithm \cite{Glandt84,gutierrez2015static,Ninarello17,Brito18}.
	The pair interaction is given by
	\begin{equation}
		U(r_{ij}) = \epsilon\left[ \left(\frac{\sigma_{ij}}{r_{ij}}\right)^{10} + \sum_{l=0}^{3} c_{2l}\left(\frac{r_{ij}}{\sigma_{ij}}\right)^{2l}\right]
	\end{equation}
	for $r_{ij}\!<\!x_c\! =\!1.4$. We use a non-additive particle size $\sigma_{ij}\!=\!\frac{1}{2}(\sigma_i+\sigma_j)(1\!-\!0.1|\sigma_i\!-\!\sigma_j|)$ to prevent crystallization and set $c_{2l}$ to enforce continuity at $x_c$ up to three derivatives. 
	We studied systems of $\mathcal{N}\!=\!2000$ particles of mass $m$ at number density $\rho = 0.58$ in cubic simulation boxes with periodic boundary conditions. We express mass in units of $m$, temperature  in units of $\epsilon$,  lengths in units $\rho^{-1/3}$, and time in units of $\sqrt{m\sigma_{\rm min}^2/\epsilon}$. 
	We minimize the energy of configurations equilibrated at temperature $T$ to produce ISs that we investigate with ASEER.\\
	\\
	For this model system, in a previous work~\cite{Massimo23}, we have (i) investigated the relaxation dynamics and shown that the self-scattering correlation function evaluated at the first peak of the static structure factor, a self-overlap function, and a total-overlap function, give consistent measures for the temperature dependence of the relaxation time $\tau$. All correlation functions satisfy the time-temperature superposition principle, which we exploit to measure the relaxation time at very low temperatures.
	(ii) estimated the microscopic time $\tau_0$ influencing the relaxation time, $\tau = \tau_0 e^{E_a(T)/T}$.\\ 
	\\
	The evaluation of $\tau$ and of $\tau_0$ allows us to measure the activation energy regulating structural relaxation, $E_a(T) = T\log(\tau/\tau_0)$. $E_a(T)$ fixes the scale of the activation energy of the excitations that have a non-negligible probability of being activated during the relaxation dynamics. The ASEER algorithm crucially allows us to extract excitations with activation energy up and beyond $E_a$.

\subsection*{Consistency between SEER and ASEER}
\begin{figure}[hbt!]
	\centering
	\includegraphics[width=0.6\linewidth]{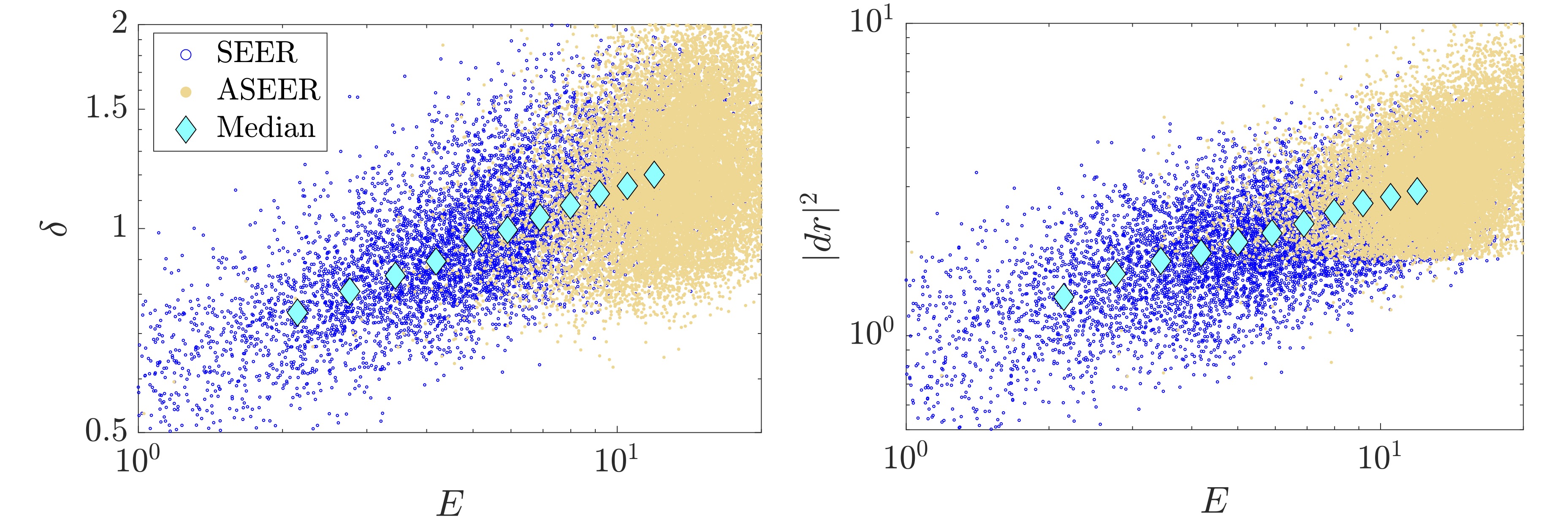}
	\caption{
		We have shown in Fig. 2a that SEER and ASEER give consistent results for the density of excitations $N(E)$ in the energy range they both access. 
		The $E$ dependence of $\delta$ and $|dr|^2$ further confirms the consistency of these methods. We also illustrate median values (in cyan) at different $E$ up to $E_a$, as shown in Fig.~4 in the main text. Data refer to $T=0.4$.} 
	\label{fig:S1}
\end{figure}
~\newpage

\subsection*{Median and mean values satisfy the scaling relations}
\begin{figure}[hbt!]
	\centering
	\includegraphics[width=0.68\linewidth]{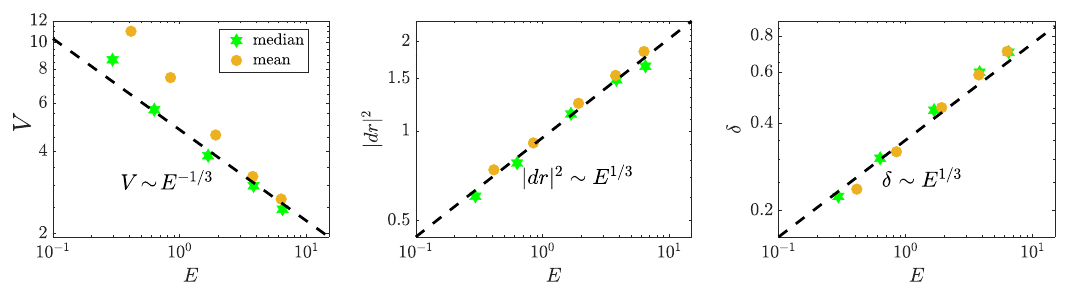}
	\caption{ 
		In the main text, we have verified the scaling relations Eq.~3 by focusing on the median values of $V\!=\!NP_r$, $|dr|^2$, $\delta$, for the lowest energy excitations across 1000 samples.
		We show in these figures that the mean values behave analogously.
	} 
	\label{fig:S2}
\end{figure}

\subsection*{Robustness of Eq.~4 with respect to the definition of $E_{\rm min}$}
\begin{figure}[hbt!]
	\centering
	\includegraphics[width=0.7\linewidth]{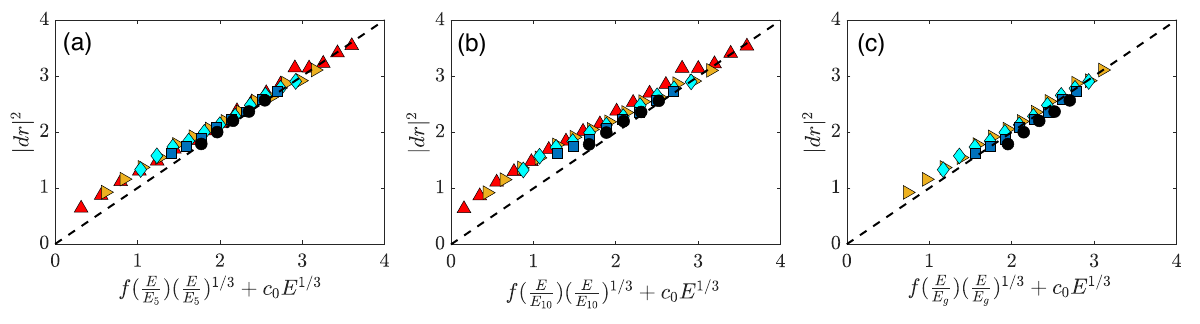}
	\caption{ 
		We validated our proposed functional form (Eq.~4) for $|dr|^2(T,E)=|dr|^2(E_{\min}(T),E)$ by adopting for $E_{\rm min}$ the median value of the lowest-energy excitations across 1000 independent samples.
		We prove the robustness of our prediction by showing that a data collapse is also obtained by replacing $E_{\rm min}$ with the median of the five lowest excitations $E_5$ (a), the median of the ten lowest excitations $E_{10}$ (b), or the gap energy $E_g$ (c).} 
	\label{fig:S3}
\end{figure}
~\newpage

\subsection*{Architecture of high-energy excitations: the role of string-like excitations }
Fig. 4 in the main text investigated the architecture of high-energy excitations that are not string-like. Including strings as well leave our results unchanged at small and intermediate energies. At very large energies, strings dominate and corrections are apparent for $V$ and $\delta$, and less so $|dr|^2$. 
\begin{figure}[hbt!]
	\centering
	\includegraphics[width=0.7\linewidth]{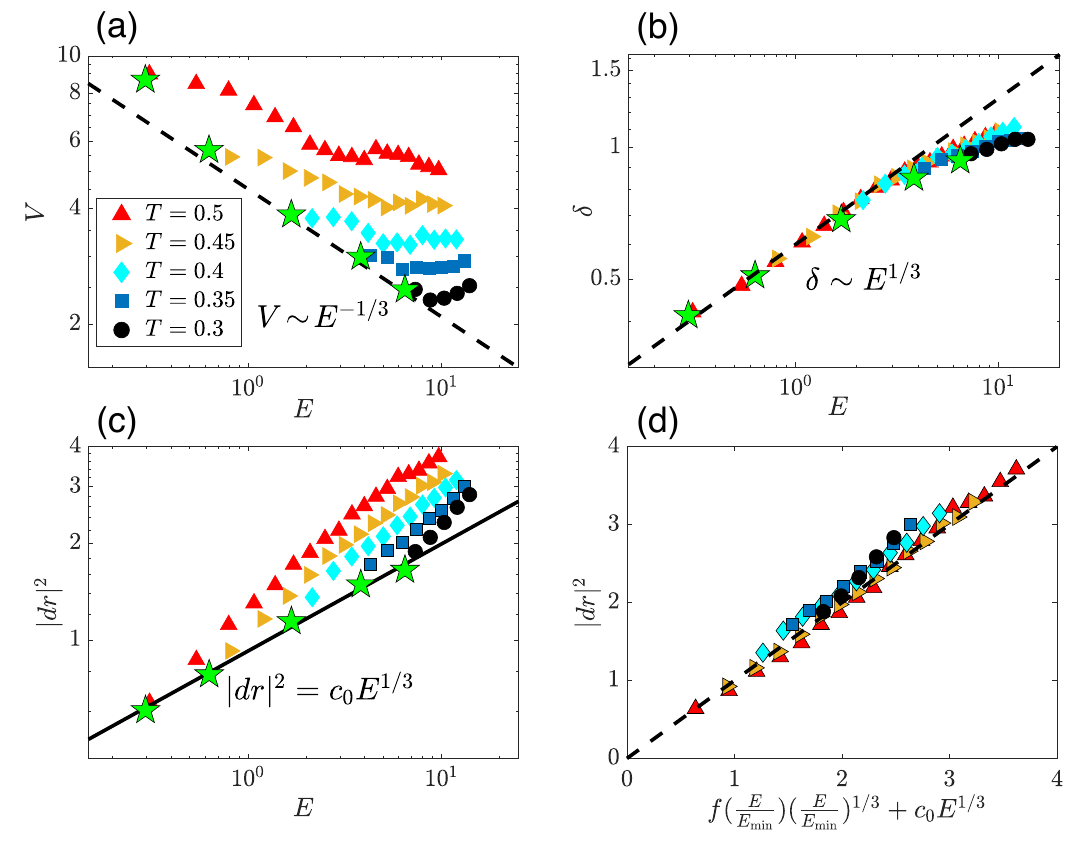}
	\caption{ 
		The analysis of the geometrical properties of non-string-like excitations illustrated in Fig. 4 in the main text is here repeated by also considering them. } 
	\label{fig:S4}
\end{figure}

\end{document}